\documentclass[aps,pre,preprint,nofootinbib,eqsecnum,tightenlines,
showpacs,amsmath,titlepage]{revtex4}
\usepackage{graphicx}
\begin{document}

\vspace*{2cm}

\title{Casimir force and its relation to surface tension}

\author{J. S. H{\o}ye}\email{johan.hoye@ntnu.no}

\affiliation{Department of Physics, Norwegian University of Science and
Technology, N-7491 Trondheim, Norway}

\author{I. Brevik}\email{iver.h.brevik@ntnu.no}

\affiliation{Department of Energy and Process Engineering, Norwegian University of Science and Technology, N-7491 Trondheim, Norway}

\date{\today}

\begin{abstract}
From energy considerations there  is reason to expect that the work done by Casimir forces during a slow displacement of the parallel plates reflects the free energy of the surface tension of the adjacent surfaces. We show this explicitly, for a one-component ionic fluid or plasma with $q_c$ as ionic charge, where the particles are neutralized by a uniform continuous oppositely charged background. For two equal half-planes, the surface-associated free energy for one half-plane turns out to be just one half of the total Casimir energy for the conventional Casimir setup. We also comment, from a wider perspective,  on the intriguing possibility that knowledge about the magnitude of the surface tension coefficient obtained from statistical mechanics or experiments may give insight into the value of the conventional cutoff time-splitting parameter $\tau=t-t'$ occurring  in quantum field theory. A simple analysis suggests that the minimal distance $\tau c$ is of the order of atomic dimensions, which is a physically natural result.

\end{abstract}

%\pacs{05.20.Jj, 11.10.Wk, 12.20.-m, 71.10-w}

\maketitle

%111111111111111111111111111111111111111111111111111111111111111111111111111111111111111111111111111111
\section{Introduction}
\label{sec1}

As is known, there are many facets of the Casimir effect: the standard transverse force between two parallel half-spaces (for reviews, see \cite{ milton01,bordag09}),  the issue concerning the temperature correction to the force (still unresolved \cite{decca05,brevik06,brevik14}),    the Casimir friction force occurring when one plate slides against the other with constant or variable velocity \cite{hoye93,pendry97,pendry10,volokitin07,barton11,hoye13,silveirinha14,hoye15} (a recent review given in \cite{milton16}), the complications that arise  if  the system is at thermal non-equilibrium \cite{antezza04,antezza05,antezza06,antezza08}, and so on. In the present paper we will focus on one aspect of the problem complex that has to our knowledge not received much attention so far, namely the association of the Casimir free energy with the {\it surface tension} of the plane surfaces. A reason to make such an association is energy balance. At large separation the two adjacent surfaces of the half-planes represent an additional energy as given by the surface tension, which is energy per unit area. This extra energy is due to particles at the surface that are less bound as they are surrounded by fewer neighboring particles. In principle the two surfaces at large separation may be created by performing work against the Casimir force. Then the initial situation is with the surfaces in contact. Physically, this is the same as one single system in bulk with no interface. Now the Casimir force can perform work between these two situations with the surfaces at infinite separation and zero separation. When this work is performed at constant temperature it is expected to be equal to the Helmholtz free energy difference.

Uniform temperature will be assumed, as well as a vacuum gap between the surfaces. Specifically, we will show this correspondence when the two half-spaces contain a neutral ionic fluid or plasma. This model prevents the Casimir energy from diverging when the two surfaces come into contact with each other. Such a simple, though physical, model thus makes one avoid the troublesome mathematical divergence that would otherwise turn up in simple Casimir theory upon material surfaces contact.  Physically, it is the Debye shielding length around charge carriers that turns out to be an important physical ingredient here. The idea of calculating the Casimir force between parallel plates on the basis of a plasma model has been presented earlier, both from a classical and a quantum mechanical point of view (in the last case using a path integral formalism) \cite{jancovici04,buenzli08,hoye09}. The statistical mechanical approach opens  new perspectives regarding the Casimir effect: instead of quantizing the electromagnetic field, one can look at the problem as one of polarizable particles that interact via the electromagnetic field. It has been shown explicitly that these two approaches are physically equivalent \cite{brevik88,hoye98,hoye01,jancovici04}. The idea has actually been made use of even in drawing connections to a Yukawa potential in a nuclear plasma \cite{ninham14}; there may be a relationship between between Casimir forces and nucleon forces mediated by mesons.

From a wider perspective, the study of the role of surface tension may be important as it points to a link between this concept and the {\it cutoff parameter} in quantum field theory. As is known, there are several cutoff parameters, but we will only consider the simple case where there is a time splitting $\tau=t-t'$ between the two spacetime points where the Green function  (or stress tensor) is evaluated. Characteristic for field theory is that the medium is regarded to be  continuous, endowed with material parameters such as permittivity and permeability, implying that  the cutoff becomes  only a mathematical parameter introduced to avoid divergences.  Now, dimensionally the cutoff can be related to surface tension. Thus, one can hope to get an idea about the magnitude of the cutoff parameter by relating it to physically-founded surface tension found in microscopic theory combined with experiments. We will briefly return to this aspect of the problem at the end of the paper.

We begin in the next section by surveying briefly the essentials of the statistical mechanical formalism, hereunder the Ornstein-Zernike equation. The potential involved in our low density Debye-H{\"u}ckel theory will be the static potential $\Phi$. Key references in this overview are \cite{hoye09} and \cite{hoye08}. In Secs. III and IV we derive the pair correlation function, and the Casimir force and free energy. In Sec. V we turn to surface considerations, showing via the internal energy Eq.~(\ref{47}) that the surface free  energy, after the infinitely-large separation contribution has been separated off, is the same as the Casimir free energy. This property has to our knowledge not been pointed out before. In Sec. VI we highlight some basic features of the entropy of the present kind of system (essentially a classical gas),  and calculate the  entropy connected with the previously calculated Casimir free energy. In Sec. VII we discuss in terms of a concrete example the mentioned possible relationship between the surface tension and the cutoff parameter in quantum field theory.

%22222222222222222222222222222222222222222222222222222222222222222222222222222222222
\section{General expressions}
\label{sec2}

Consider the generalized Ornstein-Zernike equation in statistical mechanics \cite{hoye08,ornstein14}
\begin{equation}
h({\bf r}, {\bf r}_0)=c({\bf r}, {\bf r}_0)+\int c({\bf r}, {\bf r'})\rho({\bf r'})h({\bf r'}, {\bf r}_0)d{\bf r'}. \label{2.1}
\end{equation}
where $h({\bf r}, {\bf r}_0)$ is the (pair) correlation function, $c({\bf r}, {\bf r}_0)$ the direct correlation function, and $\rho$ the particle number density. The equation above  can be taken as a definition of $c({\bf r}, {\bf r}_0)$. The generalization consists in letting the fluid be nonhomogeneous. We recall that the pair correlation function is related to the pair distribution function $g({\bf r}, {\bf r}_0)$ via
\begin{equation}
\rho({\bf r}_0)\rho({\bf r})h({\bf r}, {\bf r}_0)=g({\bf r}, {\bf r}_0)-\rho({\bf r}_0)\rho(\bf{r}). \label{2.2}
\end{equation}
 If the particles are uncorrelated,  $g=\rho({\bf r}_0)\rho({\bf r})$. For a uniform fluid, $\rho({\bf r})=$constant, $h\rightarrow h({\bf r}-{\bf r}_0)$. The function $h$ accordingly expresses the deviation from the ideal gas value.

 We will limit ourselves to weak long-range forces in the classical limit. Then, the direct correlation function is to leading order simply related to the pair interaction $\psi$    between the particles  \cite{hemmer64,lebowitz65},
 \begin{equation}
 c({\bf r}, {\bf r}_0)=-\beta \psi(|{\bf r}-{\bf r}_0|), \label{2.3}
 \end{equation}
 where as usual $\beta=1/k_BT$. This is a result following from the so-called $\gamma$ ordering, where $\gamma$ denotes the inverse range of interaction and is assumed to be small, and conforms with  the Debye-H{\"u}ckel theory for electrolytes. We consider only low densities here. For high densities the inverse Debye shielding length would be changed.

%33333333333333333333333333333333333333333333333333333333333333333333333333333333
\section{Derivation of the pair  correlation function}
\label{sec3}

Consider a one-component ionic fluid or plasma  where $q_c$ is the ionic charge. The particles are neutralized by a uniform
continuous background of opposite charge. The fluid is located in two half-planes separated by a vacuum gap of magnitude  $a$. As already mentioned, we consider the classical case of a low density plasma where Debye-H{\"u}ckel theory is valid with good accuracy. The pair correlation function $h({\bf r},{\bf r_0})$ is then determined via the electrostatic potential $\Phi$,
\begin{equation}
\nabla^2 \Phi-4\pi\beta q_c^2 \rho({\bf r})\Phi=-4\pi\delta({\bf r}-{\bf r}_0), \quad  h({\bf r},{\bf r}_0)=-\beta q_c^2 \Phi\label{20}.
\end{equation}
This follows from Eq.~(\ref{2.1}), since with Coulomb interaction $\psi=q_c^2/|{\bf r}-{\bf r_0}|$ and Eq.~(\ref{2.3}), one has $\nabla^2 c({\bf r},{\bf r_0})=4\pi\beta q_c^2\delta({\bf r}-{\bf r_0})$.

In the present case with parallel plates the particle number density is
\begin{equation}
%\begin{displaymath}
\rho({\bf r})=\left\{
\begin{array}{ll}
\rho,\quad & z<0\\
0,  \quad  & 0<z<a\\
\rho,\quad & a<z
\end{array}
\right.
\label{21}
%\end{displaymath}
\end{equation}
with equal densities $\rho={\rm const.}$ on both plates. By Fourier transform in the $x$ and $y$ directions Eq.~(\ref{20}) becomes
\begin{equation}
\left(\frac{\partial^2}{\partial z^2}-k_\perp^2-\kappa_z^2\right)\hat\Phi=-4\pi\delta(z-z_0)
\label{22}
\end{equation}
where $k_\perp^2=k_x^2+k_y^2$, the hat denoting Fourier transform. With  $\kappa^2=4\pi\beta q_ c^2 \rho$,
\begin{equation}
\kappa_z^2=\kappa^2\left\{
\begin{array}{ll}
1,\quad & z<0\\
0,  \quad  & 0<z<a\\
1,\quad & a<z.
\end{array}
\right.
\label{23}
\end{equation}
The constant $\kappa$ is the inverse Debye-H{\" u}ckel shielding length in the media. Solution of Eq.~(\ref{22}) can be written in the form
\begin{equation}
\hat\Phi=2\pi e^{q_\kappa z_0}\left\{
\begin{array}{ll}
e^{-2q_\kappa z_0}e^{q_\kappa z}/q_\kappa+Be^{q_\kappa z},\quad & z<z_0\\
e^{-q_\kappa z}/q_\kappa+Be^{q_\kappa z},\quad & z_0<z<0\\
Ce^{-qz}+C_1 e^{qz},  \quad  & 0<z<a\\
De^{-q_\kappa z},\quad & a<z
\end{array}
\right.
\label{24}
\end{equation}
where $q=k_\perp$, $q_\kappa=\sqrt{k_\perp^2+\kappa^2}$. We let $z_0$ be situated in the lower medium (thus $z_0<0$).

The electrostatic boundary conditions at $z=0$ and $z=a$ are continuous $\hat\Phi$ and $\partial\hat\Phi/\partial z$. This gives 4 equations for the unknown coefficients. By solution one may first solve for $C$ and $C_1$ in terms of $D$. This is then substituted in the other two equations to obtain the coefficients of interest
\begin{equation}
D=\frac{4qe^{(q_\kappa -q)a}}{(q_\kappa+q)^2(1-Ae^{-2qa})}, \quad A=\left(\frac{q_\kappa-q}{q_\kappa+q}\right)^2=\frac{\kappa^4}{(q_k+q)^4},
\label{25}
\end{equation}
\begin{equation}
B=B(a)=\frac{(q_\kappa-q)(1-e^{-2qa})}{q_\kappa(q_\kappa+q)(1-Ae^{-2qa})}.
\label{26}
\end{equation}

%4444444444444444444444444444444444444444444444444444444444444444444444444444444444444444444444444444444444444444444444444444
\section{Casimir force and Casimir free energy}
\label{sec4}

The Casimir force per unit area is given by Eq.~(14) in Ref.~\cite{hoye09},
\begin{equation}
f=\frac{\rho^2}{(2\pi)^2}\int\limits_{z_0<0,z>a}\hat{h}(k_\perp,z,z_0)\hat{\psi}_z'(k_\perp,z-z_0)\,dk_x dk_y dz dz_0
\label{27}
\end{equation}
where the Fourier transform $\hat h$ of the pair correlation function $h$ is
\begin{equation}
\hat h(q,z,z_0)=-\beta q_c^2 \hat\Phi=-2\pi\beta q_c^2 D e^{-q_\kappa(z-z_0)},
\label{28}
\end{equation}
and the Fourier transform $\hat \psi$ of the ionic pair interaction $\psi=q_c^2/r$ (Gaussian units) is
\begin{equation}
\hat\psi(q,z-z_0)=2\pi q_c^2 \frac{e^{-q(z-z_0)}}{q}.
\label{29}
\end{equation}
So one finds
\begin{equation}
f=-\frac{\kappa^4}{8\pi\beta}\int\limits_0^\infty
\frac{De^{-(q_\kappa+q)a}}{(q_\kappa+q)^2}q\,dq=
-\frac{1}{2\pi\beta}\int\limits_0^\infty\frac{Ae^{-2qa}}{1-Ae^{-2qa}}q^2\,dq.
\label{30}
\end{equation}
where $\hat\psi_z'=  \partial \hat{\psi} /\partial z =-q\hat\psi$ and $dk_x\,dk_y=2\pi q\,dq$. It is convenient to introduce new variables of integration
\begin{equation}
q=\kappa \sinh t,  \quad dq=\kappa \cosh{t}\,dt.
\label{30a}
\end{equation}
With this $q_\kappa=\kappa\cosh t$ and $A=e^{-4t}$, and integral (\ref{30}) becomes \cite{hoye08}
\begin{equation}
f=-\frac{\kappa^3}{2\pi\beta}\int\limits_0^\infty\frac{e^{-g(t)}}{1-e^{-g(t)}}\sinh^2 t\cosh t\, dt,
\label{31}
\end{equation}
where $g(t)=4t+2\kappa a\sinh t$.

It can be noted that in the present case the Casimir force contains only one polarization of the electromagnetic field. The reason is that our derivations are limited to the zero frequency case. Then the TM (transverse magnetic) mode reduces to the electrostatic case where only Matsubara frequency zero remains corresponding to the high temperature classical limit. Furthermore the TE (transverse electric) mode vanishes in the electrostatic case of zero frequency, and it is thus not present in expression (\ref{31}) for the force. [It can be mentioned here that this contrasts the usual Lifshitz formula for metals that is ambiguous in this respect and has lead to the controversy about the temperature correction to the Casimir force  \cite{decca05,brevik06,brevik14}.]

When the plates move the change in the Casimir free energy per unit area $F_c$ is $dF_c=-f\,da$. So with $F_c=0$ for $a=\infty$ one finds by integration
\begin{equation}
F_c=\frac{\kappa^2}{4\pi\beta}\int\limits_0^\infty \ln(1-e^{-g(t)})\sinh t \cosh t \,dt.
\label{32}
\end{equation}
When the plates are at contact, i.e.~$a=0$, one should expect that the $F_c$ outweighs the surface tension of the two surfaces at large separation. This we will investigate  in the following section.

%55555555555555555555555555555555555555555555555555555555555555555555555555555555555555555555555555555555555555555555555555
\section{Surface free energy}
\label{sec5}

There is reason to expect that the work done by the Casimir force reflects the free energy or the surface tension connected to the adjacent surfaces of the two half-planes. This requires that the free energy $F_c$ stays finite. For commonly used continuum models of dielectric media this is not the case with a diverging force when the media approach each other. To avoid this the molecular structure has to be taken into account. It is seen that  the force given by (\ref{31}) stays finite when $a\rightarrow 0$ \cite{hoye08}. (Compared to the usual diverging high temperature result the separation $a$ is replaced by $a+2/\kappa$ for large $a$.)

The task now is to perform a statistical mechanical evaluation of the free energy of the two half-planes and separate out the part due to the interaction between the two adjacent surfaces and then try to verify it to be equal to expression (\ref{32}). To do so we go via the internal energy $U$ that can be computed from the known pair correlation function $\hat{h}=-\beta q_c^2\hat{\Phi}$; cf.~(\ref{24}). So first we compute the $U_c$ that follows from the free energy (\ref{32}). And with standard thermodynamics we find
\begin{eqnarray}
\nonumber
\beta U_c&=&\beta\frac{\partial(\beta F_c)}{\partial \beta}=\kappa^2\frac{\partial(\kappa F_c)}{\partial\kappa^2}\\
&=&\frac{\kappa^2}{4\pi}\int\limits_0^\infty\left[\ln(1-e^{-g})+\kappa a\frac{\, e^{-g}}{1-e^{-g}}\sinh t\right]\sinh t \cosh t \, dt
\label{34}
\end{eqnarray}
where $g=g(t)$, $\kappa^2\sim\beta$, and $\partial\kappa/\partial \kappa^2=1/(2\kappa)$.
By partial integration of the logarithmic term one gets a term that cancels the other term to obtain
\begin{equation}
\beta U_c=-\frac{\kappa^2}{2\pi}\int\limits_0^\infty\frac{e^{-g(t)}}{1-e^{-g(t)}}\sinh^2 t\, dt.
\label{35}
\end{equation}
This is the Casimir internal energy calculated with thermodynamics from the Casimir force via the corresponding free energy; the influence from the gap $a$ contained in $g(t)=4t+2\kappa a \sinh t$.

Next we obtain the internal energy by the statistical mechanical method. To do so we can first calculate the internal energy in bulk for the {\it uniform } system. Then  the internal energy for a system of the same size with the adjacent surfaces present is found. The surface internal energy will be the difference between these two energies. Finally this is compared with the Casimir internal energy (\ref{35}) obtained from the corresponding Casimir free energy (\ref{32}).

The internal energy $U$ per unit area due to the pair interactions is (with $z$ and $z_0$ inside the half-planes)
\begin{equation}
U=\frac{\rho^2}{2(2\pi)^2}\int \hat h(k_\perp, z, z_0)\hat\psi(k_\perp, z-z_0)\,dk_x dk_y dz dz_0.
\label{36}
\end{equation}
The factor 1/2 in front prevents double counting of configurations. (As usual the very simple result for the kinetic energy of classical particles per particle $3/(2\beta)$, can be disregarded here.)

To compute the internal energy from Eq.~(\ref{36}) we split it in several contributions since the system is non-uniform consisting of two half-planes. The usual situation in fluid theory is to apply classical statistical  mechanics on  uniform systems where methods have been developed. Also the additional problem with surfaces is disregarded. However, in the present case with a low density electron gas we have been able to evaluate explicitly the pair correlation function  also in the non-uniform case.

So one contribution to the internal energy is the bulk one for uniform system. This is straightforward to obtain and goes via integral (\ref{37a}) for $L_0$ below. 	In the present case this is modified due to a surface on each half-plane. Thus the integral for $L_0$ is modified into integral (\ref{39}) for $L_1$ where the integration of $z$ is cut at the surface. In addition there is a contribution with integral (\ref{40}) for $L_2(a)$ due to the modification of the pair correlations function close to the surface. This is expressed via the coefficient $B=B(a)$ which also is influenced by the neighboring half-plane. The last contribution comes from the mutual interaction between the two half-planes expressed via the coefficient $D$ and integral (\ref{43}) for $L_3$.

In bulk the $\hat\Phi$ of Eq.~(\ref{24}) simplifies to ($z, z_0 \ll 0$)
\begin{equation}
\hat\Phi=\frac{2\pi}{q_\kappa}e^{-q_\kappa|z-z_0|}.
\label{37}
\end{equation}
So for a plane of thickness $d$ Eq.~(\ref{36}) together with (\ref{37}) and pair interaction (\ref{29}) the $z$ and $z_0$ integrations of it give the integral
\begin{equation}
L_0=\frac{1}{q_\kappa q}\int\limits_{-d}^0\int\limits_{-\infty}^\infty e^{-(q_\kappa+q)|z-z_0|}\, dzdz_0=\frac{2d}{q_\kappa q(q_\kappa+q)}
\label{37a}
\end{equation}
The limits $z=\pm\infty$ prevent surface effects. So inserting this into Eq.~(\ref{36}) with $\kappa^2=4\pi\beta q_c^2\rho$, $\hat h$ related to $\hat\Phi$ by Eq.~(\ref{20}), and pair interaction (\ref{29}) the bulk internal energy per unit area $U_b$ is ($dk_x dk_y=2\pi q\,dq$)
\begin{equation}
\beta U_b=-\frac{2\pi}{2(2\pi)^2}\left(\frac{\kappa^2}{2}\right)^2\int\limits_0^\infty L_0 q\,dq
=-\frac{\kappa^3}{8\pi}d\int\limits_0^\infty e^{-t}\,dt=-\frac{\kappa^3}{8\pi}d
\label{38}
\end{equation}
where again the new variable of integration (\ref{30a}) is used. Result (\ref{38}) is the well known one for the classical electrolyte in the Debye-H{\"u}ckel low density limit.

Now consider one half-plane with $B$ and $D$ in Eq.~(\ref{24}) neglected for simplicity. Again the half-plane is limited to a thickness $d$, but now with $z$ restricted to $-\infty<z<0$. The lower limit $-\infty$ prevents surface effects at $z_0=-d\rightarrow -\infty$ as before. But the limit $z=0$ preserves the surface effect at this position. So now we get the modified result
\begin{eqnarray}
\nonumber
L_1&=&\frac{1}{q_\kappa q}\int\limits_{-d}^0\left[\int\limits_{-\infty}^{z_0}e^{(q_\kappa+q)(z-z_0)}\,dz+\int\limits_{z_0}^0 e^{-(q_\kappa+q)(z-z_0)}\,dz\right]\,dz_0\\
&=&\frac{2d}{q_\kappa q(q_\kappa +q)}-\frac{1}{q_\kappa q(q_\kappa +q)^2}.
\label{39}
\end{eqnarray}

For the $B$-term given by Eq.~(\ref{26}) we in a similar way have
\begin{eqnarray}
\nonumber
L_2(a)=&=&\frac{B(a)}{q}\int\limits_{-d}^0\left[\int\limits_{-\infty}^{z_0}e^{(q_\kappa+q)z+(q_\kappa-q)z_0}\,dz+\int\limits_{z_0}^0 e^{(q_\kappa-q)z+(q_\kappa+q)z_0}\,dz\right]\,dz_0\\
&=&\frac{B(a)}{q_\kappa q(q_\kappa +q)}
\label{40}
\end{eqnarray}
as the two integrals turn out to be equal consistent with equal contributions from $z<z_0$ and $z_0<z$ in this case.

Clearly, when comparing with Eq.~(\ref{37a}), the first term of expression (\ref{39}) is the bulk contribution for a plane of thickness $d$ while the remaining part contributes to the surface energy. If the other half-plane is taken away, i.e.~$a\rightarrow\infty$, the whole contribution to the surface internal energy comes from
\begin{equation}
L_\infty=L_1-L_0+L_2(\infty)=-\frac{1}{q_\kappa^2(q_\kappa +q)^2}=-\frac{4q}{\kappa^4 q_k}\frac{e^{-4t}}{1-e^{-4t}}
\label{41}
\end{equation}
with $B$ given by Eq.~(\ref{26}) for $a=\infty$ and where the bulk contribution has been subtracted.

Altogether the surface internal energy per unit area $U_\infty$ for one surface will now be similar to integral (\ref{38}) with the same prefactor ($q=\kappa\sinh t$, $dq=q_\kappa\,dt$)
\begin{equation}
\beta U_\infty=-\frac{2\pi}{2(2\pi)^2}\left(\frac{\kappa^2}{2}\right)^2\int\limits_0^\infty L_\infty q\,dq
=\frac{\kappa^2}{4\pi}\int\limits_0^\infty \frac{e^{-4t}}{1-e^{-4t}}\sinh^2 t\,dt.
\label{42}
\end{equation}
This is precisely one half of minus the Casimir internal energy (\ref{35}) for $a=0$. Thus we have shown and by that can conclude that the Casimir energy can be identified with the surface energy of both surfaces taken together.

It is also of interest to check the Casimir energy against the net surface energy for finite separation $a$. Then the $D$-term is also needed.
It connects the two half-spaces so half of it with $z>z_0$ may be considered to belong to one surface while $z<z_0$ belongs to the other. Thus for one surface we have (again similar to (\ref{40}))
\begin{equation}
L_3=\frac{D}{q}\int\limits_{-\infty}^0\int\limits_{a}^{\infty}e^{(q_\kappa+q)(z-z_0)}\,dzdz_0
=\frac{D e^{-(q_\kappa+q)a}}{q(q_\kappa +q)^2}.
\label{43}
\end{equation}
With this the surface internal energy per unit area $U_a$ for separation $a$ modifies Eq.~(\ref{41}) into
\begin{equation}
L_a=L_1-L_0+ L_2(a)+L_3.
\label{43a}
\end{equation}
For the change in surface internal energy we need the difference
\begin{eqnarray}
\label{44}
\Delta L_a=L_a-L_\infty&=&L_2(a)-L_2(\infty)+L_3=\frac{E}{q_\kappa q(q_\kappa+q)^2}\\
E&=&(q_\kappa+q)[B(a)-B(\infty)]+q_\kappa De^{-(q_\kappa+q)a}
\label{44a}
\end{eqnarray}
where we recall that $B(a)$ is the coefficient $B$ for finite plane separation $a$ while $B(\infty)$ is this coefficient for infinite separation $a=\infty$.

Inserting from expressions (\ref{25}) and (\ref{26}) we find
\begin{eqnarray}
\nonumber
E&=&\left(1-\frac{q}{q_\kappa}\right)\left[\frac{1-e^{-2qa}}{1-e^{-g}}-1\right]+\frac{4q_\kappa q e^{-2qa}}{(q_\kappa+q)^2(1-e^{-g})}\\
&=&\left(1-\frac{q}{q_\kappa}\right)\frac{e^{-g}-e^{-2qa}}{1-e^{-g}}+\frac{(1-e^{-4t})e^{-2qa}}{1-e^{-g}}
=\frac{q}{q_\kappa}\frac{(1-e^{-4t})e^{-2qa}}{1-e^{-g}}
\label{45}
\end{eqnarray}
with $q=\kappa\sinh t, q_\kappa=\kappa\cosh t$, and $g=g(t)=4t+2qa$ as before ($A=e^{-4t}$). So we find
\begin{equation}
\Delta L_a=\frac{4q}{\kappa^4 q_\kappa} \frac{e^{-g}}{1-e^{-g}}
\label{46}
\end{equation}
Altogether the surface internal energy per unit area $U_a$ minus $U_\infty$ for one surface will be a straightforward extension of expression (\ref{42}) with $L_\infty$ replaced by $\Delta L_a$
\begin{equation}
\beta (U_a-U_\infty)=-\frac{\kappa^2}{4\pi}\int\limits_0^\infty \frac{e^{-g}}{1-e^{-g}}\sinh^2 t\,dt.
\label{47}
\end{equation}
Thus this surface internal energy difference for both half-planes taken together is the same as the Casimir internal energy (\ref{35}). With equal internal energies the free energies will also be the same as will follow by integration and is given by expression (\ref{32}).

%666666666666666666666666666666666666666666666666666666666666666666666666666666666666
\section{Entropy}
\label{sec6}

Entropy has been a quantity of interest and dispute in connection with Casimir interactions. Especially this has been an issue concerning the temperature dependence of the Casimir force  between metal plates. The well known Lifshitz formula turns out to be ambiguous in this respect. Depending upon how the limit of infinite dielectric constant is taken, violation of the Nernst theorem in thermodynamics has been claimed, i.e. negative entropy connected to the transverse electric (TE) field is obtained at $T=0$  \cite{brevik14,hoye16,hoye07,sushkov11,klimchitskaya12,li16}.

In view of this it can be of interest to study shortly the entropy in the present case too. However, since the classical electron gas is considered, the Nernst theorem is not an issue, and there is no TE field.

The Nernst theorem was first found and established on basis of observations. It turned out that it can be explained by the quantum mechanical nature of matter since entropy can be understood as the natural logarithm of the number of microstates times Boltzmann's constant. At $T=0$ a system is in its ground state which means just one microstate and thus zero entropy (unless degeneracy is present). With increasing temperature the number of possible microstates can only increase by which the total entropy of a system never can be negative. For classical systems the entropy usually has no lower limit when $T\rightarrow 0$.
(One may add a constant to the entropy, but this does not change the property that it has a finite lower level independent of other parameters like volume etc.~at $T=0$ from which  Nernst theorem was formulated.)

So consider the various contributions to the entropy in our case. According to thermodynamics the entropy is given by (with derivatives at constant volume)
\begin{equation}
S=\frac{1}{T}(U-F)=-\frac{\partial F}{\partial T}=k_B \beta^2\frac{\partial F}{\partial\beta}.
\label{50}
\end{equation}
This is consistent with relation (\ref{34}) between internal energy and Helmholtz free energy.

First we may consider the bulk  internal energy (\ref{38}). With relation (\ref{34}) the corresponding bulk free energy is $\beta F_b=-\kappa^3 d/(12\pi)$ as $\kappa^2\sim\beta$. The corresponding entropy is thus
\begin{equation}
S_b=-k_B\frac{\kappa^3}{24\pi}.
\label{51}
\end{equation}

The kinetic energy $(3/2)k_B T$ per particle also contributes to the entropy. For our system the contribution  $U_k$ to the internal energy per unit area will be (disregarding the uniform background).
\begin{equation}
U_k=\frac{3}{2}k_B T\rho d=\frac{3}{2\beta}\rho d.
\label{52}
\end{equation}
The corresponding free energy $F_k$ and entropy $S_k$ is then with relations (\ref{34}) and (\ref{50})
\begin{equation}
\beta F_k=\frac{3}{2}\rho d\ln \beta\,\,(+\mbox{const.}),
\label{53}
\end{equation}
\begin{equation}
S_k=-\frac{3}{2} k_B \rho d\ln\beta\,\,(+\mbox{const.}).
\label{54}
\end{equation}
(The const. term  in the entropy as well as the free energy will also contain the volume or density dependence.) Thus the classical entropy has no lower limit when $T\rightarrow0$, so Nernst theorem does not apply. It may be noted that the classical electron gas is unstable as it will prefer to have a phase transition to higher density $\rightarrow \infty$. However, real ionic particles have a hard core that prevents collapse. Thus for low temperatures there will be a phase transition to a finite density. Anyway, all the above is fully acceptable and realistic for classical systems and there is no violation of the thermodynamics for such systems.

Then consider the surface tension contribution to the internal energy $U_\infty$. With expression (\ref{42}) and $\kappa^2\sim\beta$ it follows that it is independent of temperature in the present case. As follows from relation (\ref{34}) the corresponding free energy is then $F_\infty=U_\infty$, and it is independent of temperature too and by that does not contribute to entropy. So with relation (\ref{50})
\begin{equation}
S_\infty=0.
\label{55}
\end{equation}

Finally we have the contribution to the entropy from the Casimir free energy $F_c$ as given by (\ref{32}). The corresponding internal energy is given by (\ref{34}) or (\ref{35}) which is the same as expression (\ref{47}) obtained by the statistical mechanical evaluation. With relation (\ref{50}) one can subtract free energy (\ref{32}) directly from the internal energy (\ref{34}) to obtain the Casimir entropy
\begin{equation}
S_c=k_B\frac{a\kappa^3}{4\pi}\int\limits_0^\infty\frac{e^{-g}}{1-e^{-g}}\sinh^2 t\cosh t\,dt.
\label{56}
\end{equation}
Alternatively, according to relation (\ref{50}), one can differentiate the free energy (\ref{32}) to obtain the same (since $\kappa^2\sim\beta$). One can note that this classical Casimir entropy stays positive.

%7777777777777777777777777777777777777777777777777777777777777777777777777777777777

\section{Field theory approach}
\label{sec7}

As alluded to above, the possibility of relating the surface tension - obviously a physical parameter - to the cutoff parameter in quantum field theory (QFT) is an intriguing possibility. Let us first recall how the stress tensor in QFT is constructed: one starts from the two-point function for the electromagnetic fields, where the two spacetime points $x$ and $x'$ are kept apart by a small cutoff parameter. The separation can be chosen in various ways: in the time direction, in the space direction, or a combination of both. Usually one takes the splitting  in the time direction, so that it implies a small time difference $\tau=t-t'$. We will do the same here. The purpose of  this  splitting is to avoid divergences in the final expressions of physical quantities, such as a surface stress.  After the calculation is completed, one usually omits the cutoff term, regarding it  as a mathematical artifact.  As the standard calculation of this type makes use of a complex frequency rotation, the time splitting parameter becomes proportional to the difference in imaginary time.

As we will see in a typical example, it is however possible to  obtain some insight in the physical meaning of this mathematical trick by observing the fact that the   surface tension and the time splitting parameter are related dimensionally in a simple way.

Consider a nonmagnetic dielectric ball of radius $a$, at zero temperature. The Casimir theory for it was worked out by Milton \cite{milton80}. We will look only at the limit of low susceptibility, $\varepsilon-1 << 1$, as this case is simple to handle mathematically. The surface force density was found to have the form
\begin{equation}
f=-\frac{(\varepsilon-1)^2\hbar c}{16^2 \pi a^4}\left[ \frac{16}{\delta^3}+\frac{1}{4}\right], \quad \delta =\frac{\tau c}{a},
\end{equation}
in dimensional units. Here $\delta$ is the cutoff parameter $\tau$ in nondimensional form. Both terms in the expression above are negative, corresponding to an inward directed force.  Of interest to us in the present context is the cutoff-dependent first term. Let us equate this term to the hydrodynamic surface tension stress on a compact fluid sphere of radius $a$,
\begin{equation}
\frac{(\varepsilon-1)^2}{16\pi a^4}\frac{\hbar c}{\delta^3}= \frac{2\sigma}{a}, \label{Milton}
\end{equation}
$\sigma$ denoting the surface tension coefficient. It is seen that for a ball with given permittivity  the time-splitting parameter is related to the surface tension simply as
\begin{equation}
\tau \propto \sigma^{-1/3},
\end{equation}
independently of the radius $a$.

We can also solve Eq.~(\ref{Milton}) in terms of  $\tau c$, the distance moved by a photon during the time-splitting time, to get
\begin{equation}
\tau c=6.80\times 10^{-7}\times \left[ \frac{(\varepsilon-1)^2}{\sigma}\right]^{1/3}~\rm cm.
\end{equation}
As an illustration, choose $\sigma=73~$dyne/cm, the surface tension for an air-water surface, and choose $\varepsilon-1=0.01$. Then, we get $\tau c=0.75~${\AA}, corresponding to $\tau=2.5\times 10^{-19}~$s. The important point here is that the minimum  distance $\tau c$ turns out to be {\it of the same order as atomic dimensions}.

We have to emphasize that the arguments above, indicating a  link between   microscopic statistical mechanics and field theory, are suggestive only.  One might ask if physically attractive relationships of the sort
\begin{equation}
\tau c \sim 1~{\rm \AA}, \quad \tau \sim 10^{-19}~\rm s
\end{equation}
are typical   in more general cases also. The answer to that is however not known.

%888888888888888888888888888888888888888888888888888888888888888888888
\section{Summary}
\label{sec8}

We have considered the work done by the Casimir force between parallel planes filled with a one-component ionic fluid or plasma. The ionic fluid is at low density such that the well known Debye-H{\"u}ckel theory of classical statistical mechanics for it can be applied with good accuracy. For this system we show explicitly that the work done by the Casimir force when the separation between the plates changes, reflects precisely the surface tension of the plates. A simple analysis of a corresponding quantum field theory approach suggests that its conventional time splitting parameter $\tau$ corresponds to a natural distance $\tau c$ of atomic dimensions.

\section*{Acknowledgment}

This work is supported by The Research Council of Norway, Project No. 250346.

%BBBBBBBBBBBBBBBBBBBBBBBBBBBBBBBBBBBBBBBBBBBBBBBBBBBBBBBBBBBBBBBBBBBBBBBBBBBBBBBBBBBB

\end{document}